\documentstyle[aps]{revtex}

\begin{document}
\begin{center} \begin{bf}
Analysis of the Low-Energy Theorem for
$\gamma p \rightarrow p \pi^{0}$
\end{bf} \end{center}
\begin{center}
H.W.L. Naus$^1$ and R.M. Davidson$^2$
\\
$^1${\em Institute for Theoretical Physics, University of Heidelberg} \\
{\em Philosophenweg 19, 69120 Heidelberg, Germany  }
\\
$^2${\em National Institute for Nuclear Physics and High Energy Physics}\\
{\em P.O. Box 41882, 1009 DB Amsterdam, The Netherlands}
\end{center}

\begin{abstract}
The derivation of the `classical' low-energy theorem (LET) for
$\gamma p \rightarrow p\pi^0$ is re-examined and compared to
chiral perturbation theory. Both results are correct and are not contradictory;
they differ because different expansions of the same quantity are involved.
Possible modifications of the extended partially conserved axial-vector
current relation, one of the starting points in the derivation of the LET,
are discussed. An alternate, more transparent form of the LET is presented.
\end{abstract}

The low-energy theorem (LET) for photoproduction of
neutral pions from protons is the subject of an on-going discussion.
The reason for this is that the recent calculations in the framework of
chiral perturbation theory (CHPT) \cite{ber1,ber2} and heavy baryon
chiral perturbation theory (HBCHPT) \cite{ber3} are claimed to contradict
the older result, i.e. the `classical' LET \cite{deb}. This lead to the
statement that the LET \cite{f1} is actually not a theorem
\cite{ber1,ber2,ber3,ecker}. In this Letter, we point out that the LET is a
theorem in the sense that it is based on a few general principles, and once
these are given the final result is model-independent.
It is also shown that the commonly used extrapolation \cite{deb,nfk}
of the off-shell to the physical pion-nucleon coupling can be avoided by
means of the exact, i.e. non-extrapolated Goldberger-Treiman relation.
Furthermore, the validity of the particular form of chiral symmetry
breaking used in deriving
the LET is investigated. However, as will be discussed below, the apparent
discrepancy between the CHPT \cite{f2} calculation to order $q^3$ and the LET
is due to the fact that expansions in different parameters,
pion mass versus energy, are made. Therefore, both results are correct
within their respective frameworks.

A model-independent result based on a few general principles should be
regarded as a theorem. The LET under consideration is based on Lorentz
invariance, gauge invariance, crossing symmetry, the partially conserved
axial-vector current (PCAC) {\it hypothesis}, and its extension to include
the electromagnetic interaction \cite{adler}. This hypothesis formulates the
underlying chiral symmetry and its breaking. Note that, besides spontaneous,
also explicit symmetry breaking is included. Recently, the LET  has been
carefully rederived \cite{nfk} starting from the principles mentioned above.
The consequences of isospin symmetry breaking \cite{naus1} and the explicitly
broken chiral symmetry for the LET \cite{naus2} have also been addressed.
The LET was shown not to be modified. Let us recall some details relevant for
this work, which mainly concern the implementation of extended PCAC,
\begin{equation}
(i\partial^{\mu}+e_{\pi}A^{\mu})J_{5,\mu}^{\pm ,0} =
if_{\pi}M_{\pi}^{2}\phi^{\pm ,0} \; .
\end{equation}
Here, $A_{\mu}$ is the photon field, $J_{5,\mu}$ is the axial-vector current,
$\phi$ is the pion (interpolating) field, $M_{\pi}$ is the pion mass, $f_{\pi}$
is the pion decay constant, and $e_{\pi}$ is the pion charge.
Taking the appropriate matrix element of Eq. (1), one finds
\begin{equation}
\frac{f_{\pi}M_{\pi}^{2}}{M_{\pi}^{2}-q^{2}} \langle
N(\vec{p}^{\prime})|j_{\pi}^{\pm ,0}| N(\vec{p}),\gamma (\vec{k})
\rangle = iq^{\mu} \langle N(\vec{p}^{\prime})|
J_{5,\mu}^{\pm ,0}|N(\vec{p}),\gamma (\vec{k}) \rangle
-e_{\pi} \langle N(\vec{p}^{\prime})|
J_{5,\mu}^{\pm ,0}A^{\mu}|N(\vec{p}),\gamma (\vec{k}) \rangle \; .
\end{equation}
The left-hand side of this equation contains the pion production
amplitude expressed by means of the pion source, $j_{\pi}$.
The second term on the right-hand side of Eq. (2) does not contribute
to neutral pion production. The first term on the right-hand side of
Eq. (2) requires a careful treatment because of possible nucleon pole
contributions. Therefore, the amplitude is first divided into two general
classes: class A diagrams, which contain the dressed nucleon propagator
and half off-shell $\gamma NN$ and $\pi NN$ vertices, and generalized
non-pole contributions (class B diagrams). As the above mentioned principles
provide enough constraints on these contributions to determine the LET,
no (microscopic) model or theory for the hadron structure is needed.
This does not imply, however, that the internal
structure of the hadrons is ignored.

Since Eq. (2) is a priori defined only for virtual pions \cite{nfk},
the production of off-shell pions with $\vec{q}=0$ (the pion three momentum
in the cm frame) and $q_{0}$ $\rightarrow$ 0 (the pion energy in the
cm frame), has been considered. This does not mean that the pion mass
is taken to be zero. As a technical tool, an artificial mass difference
between the in- and outgoing nucleons is introduced \cite{deb} (this
procedure is not unique and other methods, yielding the same final result,
have been used \cite{bern}). In this way, two small parameters appear,
$ \omega =q_{0}/M $ \cite{f3} and $ \delta = (M'-M)/M$. These mathematical
tools
are necessary to determine unknown amplitudes. In particular, the mass
splitting
\cite{f4} enables one to separate out possible pole terms, i.e. contributions
of order 1/$q_{0}$. After this separation, it is assumed that an expansion
of the amplitude in $\omega$ and $\delta$ is valid, with $\delta$ taking on its
physical value in the end. The expansion in $\omega$ is supposed to
hold for $q_{0}$ $<$ $M_{\pi}$ (the validity of this assumption and the
limit $q_{0}$ $\rightarrow$ $M_{\pi}$ will be discussed below).

Applying the above principles then leads to the final result, the LET for
$\gamma p \rightarrow p\pi^0$
\begin{equation}
E_{0+} = -\frac{eg}{8\pi M} \left[ \omega -\frac{\omega^2}{2}
(3+\kappa_{p}) \right] + {\cal O}(\omega^{3}) \; ,
\end{equation}
where $E_{0+}$ is the s-wave multipole.
It is evidently an expansion in the kinematical variable $\omega$, and only
contains observable hadron properties like the nucleon mass, $M$, the proton
charge, $e$, the proton
anomalous magnetic moment, $\kappa_p$, and the pion-nucleon coupling constant,
$g$. In order to compare with experiment, the limit $\omega$ $\rightarrow$
$M_{\pi}/M$ $\equiv$ $\mu$ is normally taken in Eq. (3), and the LET is usually
presented at the point $\omega$ = $\mu$. However, to exploit the main
power of the LET, i.e. a theoretical check on models,
the above limit is not required \cite{dav}.

In CHPT, one makes the plausible assumption
\cite{leut} that the low-energy regime of QCD is described by an effective
Lagrangian \cite{gass} incorporating the global symmetries of QCD. One then
arrives at a systematic expansion scheme in terms of small momenta and
meson masses, denoted by $q^n$. Up to order $q^3$, the CHPT result
\cite{ber1,ber2,ber3} is
\begin{equation}
E_{0+}= -\frac{eg}{8\pi M} \left[ \mu -\frac{\mu^2}{2}
(3+\kappa_{p}+\frac{M^{2}}{8f_{\pi}^{2}}) \right]
+ {\cal O}(\mu^{3}) \; .
\end{equation}
It should be remarked that Eq. (4) is valid at threshold, and that no
distinction is made between the pion energy and the pion mass. In other words,
the CHPT amplitude is expanded in the pion mass regardless of its origin. It is
obvious that Eqs. (3) and (4) disagree at $\omega$ = $\mu$; numerically, Eq.
(3)
gives -2.3 $\times$ 10$^{-3}M_{\pi^{+}}^{-1}$, while Eq. (4) gives +0.9 in the
same units. The current experimental result \cite{bern,berg,drech}
is -2.0 $\pm$ 0.2, and new experiments are underway at SAL and Mainz.

On the theoretical side, we do not object to the
conclusive remark in \cite{ecker}, but we propose to extend it to include the
LET: for models that satisfy extended PCAC, CHPT {\it and} the LET \begin{it}
simultaneously \end{it} serve as important theoretical checks. Nevertheless,
we do not agree with the criticisms in Ref. \cite{ecker}, which state that
several assumptions were used in the derivation of the LET which do not
hold in the `standard model'. The reason is that these assumptions
were not made in the derivation of the LET, and
we now successively address them.
First, we stress that all loop contributions
are implicitly taken into account in the derivation of the LET.
The general vertices and propagators, as well as the
general non-pole contributions, in principle contain loops, and no
contribution is dropped or neglected.

Secondly, it is often stated \cite{ber1,ber2,ber3,ecker} that one assumes
that the coefficients of the $\omega$
expansion are analytic in $\mu$, and as this is not true
in CHPT, the LET is wrong.
The divergence of some coefficients in the chiral limit
is nothing else than the Li-Pagels
mechanism \cite{li}. Already before the CHPT calculations, this potential
problem was addressed in detail, and
it was shown that the LET does not change due to this \cite{naus2}.
It was also anticipated that the coefficients of the $\mu$
expansion could be different than the coefficients of the $\omega$
expansion due to the Li-Pagels mechanism.
Differentiating between pion mass and energy, and only expanding
in the latter, avoids this mechanism and the LET is not
affected.
Thus, the assumption stated above has not been made, and the LET is
valid even if the coefficients are nonanalytic in $\mu$.
We emphasize once more that an expansion has been made in the variables
$\omega$ and $\delta$. After implementation of PCAC, the fictitious mass
difference, $\delta$ can be put to zero and one is left with an expansion
in the energy, $\omega$. Usually, however, the on-shell limit,
$\omega$ = $\mu$, is taken, and the result takes on
the {\it form} of an expansion in the pion-nucleon mass ratio. As already
recognized by Kroll and Ruderman \cite{kr} and later by Vainsthein and
Zakharov \cite{vz}, there is no a priori reason that this expression
should coincide with the $\mu$ expansion of the amplitude. The coefficients
may also depend on $\mu$ and this dependence is not constrained by the
principles used in the derivation of the LET. We stress that
the validity of the $\omega$ expansion for $\omega$
$<$ $\mu$ is not disproved in the CHPT calculation. In fact, it was
argued \cite{sfk} that the expansion converges for 0 $<$ $\omega$
$\leq$ $\mu$. Morever, it was shown \cite{sfk} that
expansion of the CHPT amplitude \cite{ber2} in terms of $\omega$
produces the LET, Eq. (3).
In summary, QCD does not forbid an energy expansion with
finite pion mass, and, as in Compton scattering, the coefficients of this
expansion may be nonanalytic in the pion mass.
We cannot, however, prove
PCAC and its extension starting from QCD.

The arguments raised above can be explicitly checked in
the linear sigma model \cite{dav} or from the CHPT amplitude \cite{sfk}.
In particular, one can test
if the assumed power expansion in $\omega$ is valid for $\omega$
$\leq$ $\mu$. It was found for both the CHPT amplitude \cite{sfk}
and the linear sigma model amplitude \cite{dav} that the expansion
converges for $\omega$ $\leq$ $\mu$, and
the coefficients of $\omega^n$
(for $n$ $\geq$ 3) are $\sim$ 1/$\mu^{(n-2)}$, in agreement with
the anticipation of Li and Pagels \cite{li}. It should be emphasized
that to obtain the LET, only an expansion in $\omega$ is needed.
In contrast to the claim in \cite{ecker},
no additional expansion in $\mu$ is needed. In order to illustrate
the behavior of the $\omega$ expansion at $\omega$ = $\mu$, the
$\omega$ coefficients were expanded in $\mu$ \cite{dav},
and it was demonstrated
that an infinite series had to be summed to obtain the CHPT result.
However, as the series converges at $\omega$ = $\mu$, this does not
imply that an `illicit interchange of limits' \cite{ecker} has been made.
The off-shell behavior of the CHPT amplitude could, of course,
be different than the off-shell behavior of the linear sigma model amplitude.
In an effective Lagrangian approach, such as
CHPT, off-shell ambiguities may arise due to the presence of terms
which vanish on-shell, see, e.g. \cite{sch1}.
We re-emphasize that starting with the above principles, the LET
is valid in the region $\omega$ $\leq$ $\mu$ and it has no
off-shell ambiguity, i.e. models that satisfy PCAC should agree on the
off-shell value of the $E_{0+}$ up to and including order $\omega^2$.

Turning to the data, the experiment happens to agree with the
numerical value of the LET, and, consequently, not with
the order $q^3$ CHPT result.
The disagreement between experiment and the CHPT result is believed
to be due to the slow convergence of the expansion in $\mu$,
and it is concluded
that this reaction is not the ideal place to test the standard model
\cite{ecker}.
Although convergence issues are beyond the predictive power of LET's in
general, one can look at convergence questions given a reasonable model.
Indeed, the $\omega$ expansion of the linear sigma
model amplitude \cite{dav}
converges slowly as $\omega$ $\rightarrow$ $\mu$, and therefore
the agreement of the LET with the data is somewhat surprising.
Another problem in CHPT is that the isospin
violation corrections are not fully understood, but expected to be large.
In contrast, the LET for this reaction is only trivially modified \cite{naus1}
by isospin violation corrections.
In the linear sigma model, the main source of difference between
the exact model result and the LET value arises from the intermediate
$n\pi^+$ state. In nature, this threshold is 6 MeV above the
$p\pi^0$ threshold. Including isospin symmetry breaking (by hand)
in the linear sigma model, it was found
that the LET value is more accurate, but still 35\% higher than the
exact model result \cite{dav}.

The result of the derivation in Ref. \cite{nfk} actually contains
the off-shell pion-nucleon coupling at $q^2=0$, i.e. $g(0)$.
Of course, given the order of $\omega$, it is consistent to replace it by
$g(q_{0}^{2})$. Finally putting in $\omega$ = $\mu$ yields
the LET in terms of the physical pion-nucleon constant, $g$.
This procedure appears like the extrapolation usually made in the
derivation of the Goldberger-Treiman relation.
In fact, by use of the exact Goldberger-Treiman relation, i.e.
{\it not} using the extrapolation, one can immediately express the result
in terms of physical quantities.
To demonstrate this explicitly, recall that PCAC yields
a relation between $g(0)$ and the physical
weak interaction constants $g_A$, the nucleon axial-vector coupling
constant, and $f_{\pi}$,
\begin{equation}
g_{A}= \cos \theta_{c} \frac{f_{\pi}g(0)}{M} \; ,
\end{equation}
where $\theta_c$ is the Cabibbo angle.
Given the PCAC hypothesis (one of the basic ingredients in the
derivation) this is exact. What is commonly known as the
Goldberger-Treiman relation
follows from the extrapolation mentioned above, i.e., $g(0)$ $\approx$
$g(M_{\pi}^{2})$.
However, the LET involves $g(0)$ \cite{nfk}, and thus we can use
the exact relation, Eq. (5). The final expression then reads
\begin{equation}
E_{0^{+}}= -\frac{eg_{A}}{8\pi f_{\pi}\cos \theta_{c}} \left[ \omega
-\frac{\omega^2}{2}
(3+\kappa_{p}) \right] + {\cal O}(\omega^{3}) \; .
\end{equation}

In the discussions above, we have refuted the criticisms of the LET given
in Refs. \cite{ber1,ber2,ber3,ecker} and re-established the LET
based on extended PCAC.
We note here that the extension of the PCAC relation
was derived assuming minimal electromagnetic coupling\cite{adler}.
In effective theories,
however, non-minimal terms can be present.
The pertinent question is whether these terms would change the
extended PCAC relation. Consequently, the low-energy expansion
could be modified by
a contribution of the form $q_{0}M_{\pi}^{2}$, for instance.
As an explicit example we consider CHPT. Although
such a contribution is not present in the order $q^3$ calculation
of \cite{ber2}, it does appear at order $q^4$. Explicitly, one has,
in the notation of \cite{krause}, the non-minimal term \cite{koch}
\begin{equation}
{\cal L} = ia \; Tr \left[ \bar{B} \gamma_{5} \sigma_{\mu \nu}
(F^{+\mu\nu}\rho +\rho F^{+\mu\nu})B \right] \; ,
\end{equation}
where $a$ is an unknown constant.
Looking at the neutral pion-nucleon sector, we find
\begin{equation}
{\cal L} \sim eaM_{u}F^{\mu\nu} \left[ 2\bar{p}\gamma_{5}
\sigma_{\mu\nu}p\pi^{0}+\bar{n}\gamma_{5}\sigma_{\mu\nu}
n\pi^{0} \right] \; .
\end{equation}
The contribution to the $E_{0^{+}}$ at the tree level is $\sim$
$q_{0}M_{u}$ $\sim$ $q_{0}M_{\pi}^{2}$.
As anticipated, this term, Eq. (7), also gives a
contribution to the divergence of the axial-vector current,
\begin{equation}
\Delta\partial_{\mu}J_{5,0}^{\mu} \sim eaM_{u}F^{\mu\nu}
\left[ 2\bar{p}\gamma_{5}\sigma_{\mu\nu}p+\bar{n}\gamma_{5}
\sigma_{\mu\nu}n \right] \; .
\end{equation}
Therefore, Eqs. (1) and (2) are modified in CHPT, obviously leading to
the possibility
of contributions $\sim$ $\omega\mu^{2}$ in Eq. (3) \cite{ulf}.
This contribution
is evidently supressed compared to the linear term in Eq. (3).

It is important to recall that in theories with minimal electromagnetic
coupling the extension of PCAC(-like) relations is known \cite{adler}.
For example, in QCD, the divergence of the axial-vector current is
expressed in terms of quark fields. The resulting expression is an
acceptable pion interpolating field \cite{Lee},
$\phi^{\pm ,0} \sim \bar{q} \gamma_{5} \tau^{\pm 0} q $ \cite{latt}.
This choice of the pion interpolating field immediately yields PCAC;
since this cannot be proved, PCAC remains a hypothesis \cite{Lee}.
Electromagnetic coupling to the elementary quarks leads to the extended
PCAC relation. In other words, given the PCAC hypothesis starting from QCD,
corresponding to the above choice of the
interpolating pion field, its extension holds.
Note, however, that the choice of the pion
interpolating field in terms of quarks fields is not unique \cite{latt}.
Other choices modify the PCAC relation and it would be interesting to
study the possibilities and consequences of such modifications. In CHPT,
for instance, PCAC is most likely modified at some order.

Probably the best known example of a LET is Compton scattering
\cite{low,gell}. From the theoretical point of view, there is no
discussion about this LET.
On the other hand, before the connection with experiment can
be made, a careful treatment of infrared divergences is needed \cite{bj}.
In one of the two classical derivations, a discussion on
infrared divergences is included. Low \cite{low} explicitly states that the
proof applies to all orders in the electromagnetic coupling,
provided the virtual photons
are given a fictitious mass $\lambda$. He cannot guarantee its validity
in the limit $ \lambda \rightarrow 0$.
Surprisingly, the situation  for neutral pion photoproduction is
under better control. First, infrared divergences concerning virtual
photons are not present because the LET is only valid to first order
in the electromagnetic coupling. Secondly, there is no need to
give the virtual pions a fictitious mass because they are already massive.
A few clarifying remarks are in order. The fictitious mass difference
for the nucleons was introduced to deal with the nucleon pole in the
matrix element of the axial-vector current. This is not connected with
infrared problems; moreover, in the end we can take the equal mass limit.
Finally, the energy expansion holds from zero up to the pion threshold.
Beyond this threshold one cannot make definite statements.

In summary,
the `classical' LET derived {\it assuming} the extended PCAC relation
has been verified.
In particular, the LET does not break down due to neglect
of loops, as they are implicitly taken into account,
nor due to assumptions about the expansion in the pion energy.
It has been proposed to present the LET in the form of an energy
expansion, reflecting its real content. Moreover, it was demonstrated that
the extrapolation of the pion-nucleon coupling can be avoided by
using the exact Goldberger-Treiman relation.
Furthermore, we pointed
out that non-minimal contact couplings, often appearing in effective
theories, may effect the extended PCAC relation.
This was explicitly shown in CHPT, and the consequences for
its low-energy expansion were exhibited.
We also commented on the PCAC hypothesis in the context of QCD,
where, given PCAC, its extension follows.
Finally, a brief discussion of issues related to the
expansion of the Compton amplitude and its similarities to the
expansion used in deriving the LET was presented.

\bigskip
\baselineskip=12pt
H.W.L.N. is supported in part by the Federal Ministry of Research and
Technology (BMFT) under contract number 06 HD 729.
R.M.D. is supported by the
Foundation for Fundamental Research on Matter (FOM) and the National
Organization for Scientific Research (NWO), The Netherlands.


\begin{thebibliography}{99}
\bibitem{ber1}V. Bernard, N. Kaiser, J. Gasser, and U.-G.
Mei{\ss}ner, Phys. Lett. B {\bf 268}, 291 (1991).
\bibitem{ber2}V. Bernard, N. Kaiser, and U.-G. Mei{\ss}ner,
Nucl. Phys. {\bf B383}, 442 (1992).
\bibitem{ber3}V. Bernard, N. Kaiser, J. Kambor, and U.-G.
Mei{\ss}ner, Nucl. Phys. B {\bf 388}, 315 (1992).
\bibitem{deb}P. de Baenst, Nucl. Phys. {\bf B24}, 633 (1970).
\bibitem{f1}From now on we omit the word `classical'.
\bibitem{ecker}G. Ecker and Ulf-G. Mei{\ss}ner, Univ. Wien
preprint UWThPh-1994-93, (1994).
\bibitem{nfk}H.W.L. Naus, J.H. Koch, and J.L. Friar, Phys. Rev. C
{\bf 41}, 2852 (1990).
\bibitem{f2}No distinction between CHPT and HBCHPT will be made in this Letter.
\bibitem{adler}S.L. Adler, Phys. Rev. {\bf 139}, B1638 (1965).
\bibitem{naus1}H.W.L. Naus, Phys. Rev. C {\bf 43}, R365 (1991).
\bibitem{naus2}H.W.L. Naus, Phys. Rev. C {\bf 44}, 531 (1991).
\bibitem{bern} A.M. Bernstein and B.R. Holstein, Comments Nucl. Part.
Phys. {\bf 20}, 197 (1991).
\bibitem{f3}In Ref. \protect\cite{nfk}, $\omega$ was denoted by $\epsilon$.
\bibitem{f4}After emission of the pion, the nucleon has mass $M^{\prime}$.
In this way, no problems with gauge invariance occur.
\bibitem{dav}R.M. Davidson, Phys. Rev. C {\bf 47}, 2492 (1993).
\bibitem{leut} H. Leutwyler, Lectures given at the XXX Internationale
Universit\"{a}tswochen f\"{u}r Kernphysik, Schladming, Austria and
at the advanced Theoretical Study Institute in Elementary Particle
Physics, Boulder, Colorado (1991).
\bibitem{gass} J. Gasser, M.E. Sainio and A. \v{S}varc, Nucl. Phys. {\bf B307},
779 (1988).
\bibitem{berg} J. Bergstrom, Phys. Rev. C {\bf 44}, 1768 (1991).
\bibitem{drech} D. Drechsel and L. Tiator, J. Phys. G: Nucl. Part. Phys.
{\bf 18}, 449 (1992).
\bibitem{li}L.F. Li and H. Pagels, Phys. Rev. Lett. {\bf 26},
1204 (1971).
\bibitem{kr} N.M. Kroll and M.A. Ruderman, Phys. Rev. {\bf 93}, 233 (1954).
\bibitem{vz} A.I. Vainsthein and V.I. Zakharov, Sov. J. Nucl. Phys.
{\bf 12}, 333 (1971);  Nucl. Phys. {\bf B36}, 589 (1972).
\bibitem{sfk}S. Scherer, J.H. Koch, and J.L. Friar, Nucl. Phys.
{\bf A552}, 515 (1993).
\bibitem{sch1} S. Scherer and H.W. Fearing,
preprint TRI-PP-94-64 (1994); Phys. Rev. C {\bf 51}, 359 (1995).
\bibitem{krause}A. Krause, Helv. Phys. Acta, {\bf 63}, 3 (1990).
\bibitem{koch}We thank Prof. J.H. Koch for bringing to our attention the
possibility of such a term. According to him, the original suggestion
of such a term is from Prof. J. Gasser.
\bibitem{ulf}Such a term also appears in a recent work by
V. Bernard, N. Kaiser, and Ulf-G. Mei{\ss}ner, preprint CRN 94-62 (1994).
However, the relevance of this term to the LET is not discussed.
\bibitem{Lee} T.D. Lee, \begin{it} Particle Physics and Introduction to
Field Theory, \end{it} (Harwood, Chur, 1988).
\bibitem{latt}Interpolating fields of this form are succesfully used
in lattice calculations. See, e.g.,
C. Rebbi, \begin{it} Lattice Gauge Theories and
Monte Carlo Simulations,\end{it} (World Scientific, Singapore, 1983), and,
J.W. Negele, in \begin{it}Hadrons and Hadronic Matter,\end{it}
NATO ASI Series B228 (D. Vautherin, F. Lenz, and J.W. Negele, eds.,
Plenum, New York, 1990) p. 369.
\bibitem{low}F.E. Low, Phys. Rev. {\bf 96}, 1428 (1954).
\bibitem{gell}M. Gell-Mann and M.L. Goldberger, Phys. Rev. {\bf 96}, 1433
(1954).
\bibitem{bj}J.D. Bjorken and S.D. Drell, \begin{it}Relativistic Quantum
Fields,\end{it} (McGraw-Hill, New York, 1965).
\end{thebibliography}
\end{document}